\documentclass[%
 reprint,
 amsmath,amssymb,
 aps,
]{revtex4-1}

\usepackage{graphicx}
\usepackage{xcolor}
\usepackage{dcolumn}
\usepackage{bm}
\usepackage{multirow}
\usepackage{ulem}
\usepackage{hyperref}

\usepackage{subfigure}
\graphicspath{{./}{figs/}}

\begin{document}

\preprint{APS/123-QED}

\title{The asymptotic coarse-graining formulation of slender-rods, bio-filaments and flagella}

\author{Cl\'{e}ment Moreau$^{1}$}
\author{Laetitia Giraldi$^{1}$}%
\author{Hermes Gad\^{e}lha$^{2}$}
\email{hermes.gadelha@york.ac.uk}
\affiliation{$^1$Universit\'{e} C\^{o}te d'Azur, Inria, CNRS, LJAD, McTAO team, France \\
$^2$Department of Mathematics, University of York, YO10 5DD, UK}%

\date{\today}

\begin{abstract}
The inertialess fluid-structure interactions of active and passive inextensible filaments and slender-rods are ubiquitous in nature, from the dynamics of semi-flexible polymers and cytoskeletal filaments to cellular mechanics and flagella. The coupling between the geometry of deformation and the physical interaction governing the dynamics of bio-filaments is complex. Governing equations negotiate elastohydrodynamical interactions with non-holonomic constraints arising from the filament inextensibility. Such elastohydrodynamic systems are structurally convoluted, prone to numerical erros, thus requiring penalization methods and high-order spatiotemporal propagators. The asymptotic coarse-graining formulation presented here exploits the momentum balance in the asymptotic limit of small rod-like elements which are integrated semi-analytically.  This greatly simplifies the elastohydrodynamic interactions and overcomes previous numerical instability. The resulting matricial system is straightforward and intuitive to implement, and allows for a fast and efficient computation, over than a hundred times faster than previous schemes. Only basic knowledge of systems of linear equations is required, and implementation achieved with any solver of choice. Generalisations for complex interaction of multiple rods, Brownian polymer dynamics, active filaments and non-local hydrodynamics are also straightforward. We demonstrate these in four examples commonly found in biological systems, including the dynamics of filaments and flagella.  Three of these systems are novel in the literature. We additionally provide a Matlab code that can be used as a basis for further generalisations.  
\end{abstract}

\pacs{Valid PACS appear here}
\maketitle


\section{\label{sec:intro}Introduction}

The fluid-structure interactions of semi-flexible filaments are found everywhere in nature \cite{book:antman,book:Alberts,howard2001}, from the mechanics of DNA strands and the movement of polymer chains to complex interaction involving cytoskeletal microtubules and actin cross-linking architectures and filament-bundles and flagella  \cite{Gadelha2010,Camalet,Bourdieu95,Goldstein1995,Fu2008,Yu,olson_modeling_2013,Tornberg2003,kantsler_fluctuations_2012,sanchez_spontaneous_2012,brangwynne_microtubules_2006,plouraboue2016identification,Heussinger2010,Claessens2008,claessens2006actin}. The elastohydrodynamics of filaments permeate different branches in mathematical sciences, physics and engineering, and their cross-fertilizing intersects with biology and chemistry. The wealth of theoretical and experimental studies on the movement of semi-flexible filaments, termed here as filaments, is extensive, thus reflecting the fundamental importance of the physical interactions marrying fluid and elastic phenomena. Hitherto the elastohydrodynamics of active and passive filaments have shed new light into bending, buckling, active matter and self-organisation, as well as bulk material properties of interacting active and passive fibres across disciplines \cite{hines_bend_1979,Gadelha2010,Camalet,Bourdieu95,Goldstein1995,Fu2008,Yu,olson_modeling_2013,Tornberg2003,kantsler_fluctuations_2012,sanchez_spontaneous_2012,coy_counterbend_2017,schoeller_flagellar_2018}.

The movement of semi-flexible filaments bridges complex fluid and elastic interactions within a hierarchy of different approximations \cite{lauga2009hydrodynamics}. Here, we focus on systems governed by low Reynolds number inertialess hydrodynamics \cite{Purcell}. Both the hydrodynamic and elastic interactions of filaments are greatly simplified by exploiting the filament slenderness \cite{lauga2009hydrodynamics,book:antman}, reducing the dynamics to effectively a one-dimensional system \cite{hines_bend_1978}. A variety of model families have been developed exploiting such slenderness property, and thus it would be a challenging task to review the wealth of theoretical and empirical developments to date here. Instead we direct the reader to excellent reviews on the subject \cite{lauga_hydrodynamics_2009,powers2010dynamics,shelley2011flapping,brennen1977fluid}.

In a nutshell, two theoretical descriptions are popularly used: the discrete and continuous formulation. In discrete models, such as the beads model, gears model, n-links model, or similarly worm-like chain models (see \cite{alouges_self-propulsion_2013,alouges2015can,GiraldiMartinon13,GiraldiMartinon15,schoeller_flagellar_2018,delmotte_general_2015,plouraboue2016identification,Heussinger2010,cosentino_lagomarsino_hydrodynamic_2005,shelley2011flapping,montenegro-johnson_spermatozoa_2015,brokaw2014computer,JohnsonBrokaw79}), the filament is broken into a discrete number of units, such as straight segments, spheres or ellipsoids. The elastic interaction coupling neighboring nodes/joints is described via constitutive energy functionals or via discrete elastic connectors encoding the filament's resistance to bending. The shape of each discrete unit defines the hydrodynamical interaction, i.e. hydrodynamics of spheres for the beads and gear model, and slender-body hydrodynamics for straight rod-like elements.  Continuous models, on the other hand, recur to partial differential equation (PDE) systems to describe the combined action from fluid-structure interactions \cite{hines_bend_1979,Tornberg2003}. The dynamics arises through the total balance of contact forces and moments along the filament \cite{book:antman}. This formalism results invariably in a nonlinear PDE system coupling a hyperdiffusive fourth-order PDE with a second-order boundary value problem (BVP) required to ensure inextensibility via Lagrange multipliers \cite{hines_bend_1979,Tornberg2003,Gadelha2010}, in addition to six boundary conditions and initial configuration for closeness. The geometrical coupling guarantees that the order of the PDE remains unchanged under transformation of variables, from the position of filament centerline $X(s,t)$ at an arclength $s$ and time $t$ relative to a fixed frame of reference, to tangent angle $\theta(s,t)$ or curvature $\kappa(s,t)$ of the filament \cite{Goldstein1995,Wiggins2}. While the equivalence between discrete and continuum models is generally not available, both theoretical frameworks suffer from numerical instability and stiffness arising from the nonlinear geometrical coupling between the filament's curvature and its inextensibility constraint \cite{klapper_biological_1996,Tornberg2003}. Nonlinearities originated from curvature are well known to drive numerical instability in moving boundary systems, as found in pattern formation of interfacial flows driven by surface tension \cite{hou_removing_1994}, as well as in elastic and fluid stresses in shells and fluid membranes \cite{hou_removing_1998, rodrigues_semi-implicit_2015}. The latter often requires numerical regularization, such as the small-scale decomposition \cite{hou_removing_1994,hou_removing_1998}.

Contact forces of inextensible filaments are not determined constitutively \cite{antmannonlinear}, and require Lagrange multipliers to ensure strict length constraints.  The resulting systems in both discrete and continuous models are thus prone to numerical instabilities  \cite{lauga_hydrodynamics_2009,powers2010dynamics,shelley2011flapping,brennen1977fluid,hines_bend_1978,montenegro-johnson_spermatozoa_2015,brokaw2014computer,JohnsonBrokaw79,klapper_biological_1996,Camalet,schoeller_flagellar_2018,delmotte_general_2015,plouraboue2016identification,Heussinger2010}. This is despite the fact that discrete models automatically satisfy the length constraint by construction \cite{lauga_hydrodynamics_2009,hines_bend_1978,brokaw2014computer,plouraboue2016identification,Camalet,schoeller_flagellar_2018,delmotte_general_2015,cosentino_lagomarsino_hydrodynamic_2005,lowe_dynamics_2003,cosentino_lagomarsino_hydrodynamic_2005}, or equivalently the tangent angle formulation $\theta(s,t)$ for continuous models \cite{Camalet,hines_bend_1978,wiggins1998trapping,young_hydrodynamic_2009}, which intrinsically preserves lengths by definition. In continuum models, penalization strategies are required to regularize length errors that vary dynamically \cite{Tornberg2003,Gadelha2010,montenegro-johnson_spermatozoa_2015}. The number of boundary conditions is large, and the non-linear coupling makes complex boundary systems challenging \cite{montenegro-johnson_spermatozoa_2015}, as we discuss below. The latter imposes severe spatiotemporal discretisation constraints, increases the computational time and numerical errors, especially for deformations involving  large curvatures. 


The aim of this paper is to resolve the bottleneck arising from the interaction  between the hyperdiffusive elastohydrodynamics and the inextensibility constraint. For this, we consider a hybrid continuum-discrete approach. The coarse-graining formulation is a direct consequence of the asymptotic integration of the moment balance system along coarse-grained rod-like elements.  No explicit length constraint is required, and the resulting linear system is structurally stable and does not require explicit computation of the unknown force distribution aforementioned.  Numerical implementation is straightforward and allows for faster computation, over than a hundred times faster, with increasingly better performance for tolerance to error below 1$\%$.  This greatly decreases the implementation complexity, the number of boundary conditions required, computational time and numerical stiffness. The coarse-graining framework can be readily applied to systems that would be prohibitive using the classical system, as we discuss in section \ref{numerical_comp}. Furthermore, we show that the coarse-graining implementation is simple, and generalisations for complex interaction of multiple rods, Brownian polymer dynamics, active filaments and non-local hydrodynamics are straightforward. 

This manuscript is structured as follows: first, we describe the momentum balance for an inextensible filament embedded in an inertialess fluid, and re-derive the classical elastohydrodynamic system in Section \ref{model}. For this, we employ the standard elastic theory for slender-rods and lowest order hydrodynamic approximation for slender-bodies, i.e. Resistive Forces Theory \cite{GrayHancock55}. In Section \ref{section:coarse}, we introduce the asymptotic coarse-graining formulation. In Section \ref{numerical_comp}, we contrast the classical elastohydrodynamics and the coarse-graining formulations and their respective numerical performances. Finally, we abandon the classical  elastohydrodynamic formulation and explore several systems with  the coarse-graining approach in Section \ref{applications}. We investigate the buckling instability of a bio-filaments \cite{Bourdieu95,Tornberg2003}, magnetically-driven micro-swimmers \cite{alouges2015can,GiraldiPomet16,gutman2014simple,gadelha2013optimal}, the counterbend phenomenon for effectively one-dimensional filament-bundles \cite{gadelha_counterbend_2013,Gadelha_elastica,coy_counterbend_2017} and the driven motion of a two-filament bundle assembly. Expect from the magnetic swimmer, the other bio-filament systems are entirely novel in the literature. We also provide the Matlab code via github free repository that can be used a basis for further generalisations.

\begin{figure*}[t!]
\begin{center}
\includegraphics[width=\textwidth]{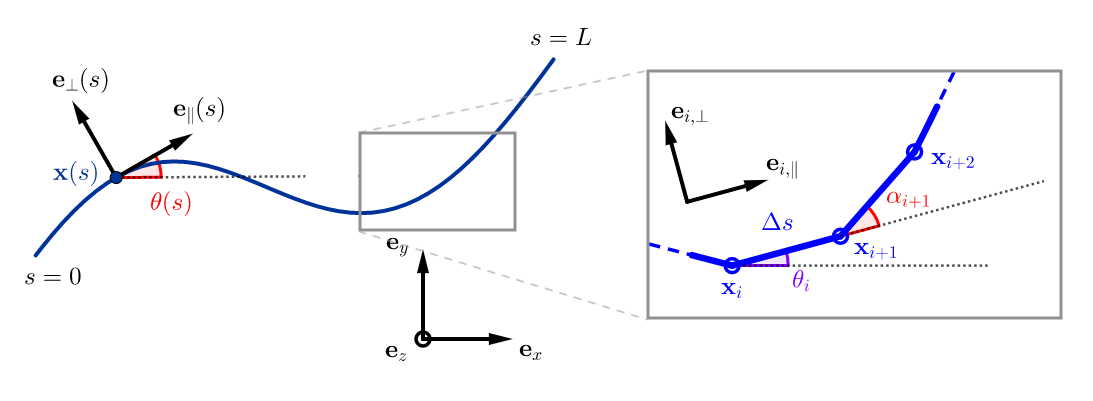}
\caption{Parametrization of the continuous and discrete filaments}
\label{schema_filament}
\end{center}
\end{figure*}

\section{Classical elastohydrodynamic filament theory \label{model}}
Consider an inextensible elastic rod of length $L$, parametrized by its arclength. The position of a point of arclength $s$ on the filament is denoted by $\mathbf{x}(s)$. The filament can experience two types of forces  \cite{book:antman}: contact forces $\mathbf{n}(s)$ within the filament, and external forces, that have a force density $\mathbf{f}(s)$ (by unit of length), later this will incorporate the hydrodynamic interaction. The second Newton's law ensures the momentum balance:
\begin{equation}
\mathbf{n}_s+\mathbf{f}=0,
\label{pde_1}
\end{equation}
\begin{equation}
\mathbf{m}_s+\mathbf{x}_s\times \mathbf{n}= 0,
\label{pde_2}
\end{equation}
where the subscripts denote derivatives with respect to arclength $s$, $\mathbf{m}(s)$ is the contact moment, and external moments are neglected. The dynamical system~\eqref{pde_1}-\eqref{pde_2} is further specified by the geometry of the deformation and the constitutive relations characterising the filament. Here we focus on inextensible, unshearable hyperelastic filaments undergoing planar deformations. Thus the contact forces are not defined constitutively whilst the bending moment is linearly related to the local curvature \cite{book:antman}.

The position of the filament centerline is denoted by $\mathbf{x}(s,t)$. The Frenet basis moving with the filament is given by $(\mathbf{e}_{\parallel},\mathbf{e}_{\bot})$, tangent and normal vector respectively. The angle between the $x$-axis of frame of reference and $\mathbf{e}_{\parallel}$ is $\theta$, where the normal vector to the plane in which deformation occurs is $\mathbf{e}_z$, see Figure \ref{schema_filament}. The filament is characterised by a bending stiffness $E_b$, and thus elastic moments are simply $\mathbf{m}(s)= E_b \theta_s \mathbf{e}_z$. The latter can be used in conjuction with~\eqref{pde_2}, using $\theta_{ss} \mathbf{e}_{\bot}=\mathbf{x}_{sss}$, to get
\begin{equation}
\mathbf{n}(s)=- E_b \mathbf{x}_{sss} \mathbf{e}_{\bot} + \tau \mathbf{e}_{\parallel},
\label{eq:n}
\end{equation}
where $\tau(s)$ is the unknown Lagrange multiplier. The hydrodynamical friction experienced by a slender-body in low Reynolds number regime can be simplified asymptotically by employing the Resistive Force Theory \cite{GrayHancock55}, in which hydrodynamic friction is related with to velocity via an anisotropic operator
\begin{equation}
\mathbf{f}(s)=-\xi (\mathbf{e}_{\bot}\cdot \mathbf{x}_{t})\mathbf{e}_{\bot} + \eta (\mathbf{e}_{\parallel}\cdot \mathbf{x}_{t}) \mathbf{e}_{\parallel}  \,,
\label{eq:fhydro}
\end{equation}
where $\eta$ and $\xi$ are the parallel and perpendicular drag coefficients, respectively. Using~\eqref{pde_1} and nondimensionalizing the system with respect to the length scale $L$, time scale $\omega^{-1}$, force density $E_b/L^3$, and noticing that $\mathbf{e}_{\parallel}=\mathbf{x}_s$, the dimensionless elatohydrodynamic equation for a passive filament deforming in a viscous environment reads:
\begin{equation}
\mathrm{Sp}^4 \mathbf{x}_t = - \mathbf{x}_{ssss}-(\gamma-1)(\mathbf{x}_s \cdot \mathbf{x}_{ssss})\mathbf{x}_s + (\tau \mathbf{x}_{ss}+ \gamma \tau_s \mathbf{x}_s),
\label{eq:pde}
\end{equation}
with the dimensionless parameters $\mathrm{Sp}=L (\omega \xi/E_b)^{1/4}$ and $\gamma=\xi/\eta$. The unknown line tension is obtained by invoking the inextensibility constraint 
\begin{equation}
\frac{\partial}{\partial t}(\mathbf{x}_s \cdot \mathbf{x}_s)=0,
\label{equation_3}
\end{equation}
which together with~\eqref{pde_1} provides a nonlinear second-order boundary value problem for the line tension, or Lagrange multiplier,
\begin{equation}
\gamma \tau_{ss}-(\mathbf{x}_{ss} \cdot \mathbf{x}_{ss})\tau = -3 \gamma (\mathbf{x}_{sss} \cdot \mathbf{x}_{sss})-(3 \gamma + 1) (\mathbf{x}_{ss} \cdot \mathbf{x}_{ssss}).
\label{eq:tension}
\end{equation}
In practice, however, this inextensibility condition is prone to numerical erros \cite{Tornberg2003} causing the filament length to vary over time. A penalisation term is thus added on the right-hand side of ~\eqref{eq:tension} to remove spurious incongruousnesses of the tangent vector \cite{Tornberg2003,Gadelha2010,montenegro-johnson_spermatozoa_2015}.

The non-linear, geometrically exact elastohydrodynamical system Eqs.~\eqref{eq:pde} and~\eqref{eq:tension} requires a set initial and boundary conditions for closeness. At the filament boundaries, either the force/torque are specified or the endpoints kinematics is imposed. Here we consider the distal end free from external forces and moments
\[
\forall t, \quad - \mathbf{x}_{sss}(L,t) + \tau \mathbf{x}_s (L,t) = 0, \quad \mathbf{x}_{ss} (L,t)=0.
\]
At the proximal end, several scenarios may be considered: (i) Free torque and force condition, thus the above equations are satisfied at $s=0$.
(ii) Pivoting, pinned or hinged condition: the extremity has a fixed position but it is free rotate around it, $\mathbf{x}_t (0,t) =0, \mathbf{x}_{ss} (0,t) =0$. (iii) Clamped condition: the extremity has a fixed position and orientation, $\mathbf{x}_t (0,t)=0, \mathbf{x}_{st} (0,t)=0$. Finally, initial conditions are required for  closeness. Boundary conditions for the Lagrange multiplier $\tau$ boundary value problem~\eqref{eq:tension} is derived from the above boundary constraints accordingly, and are generally unknown. Thus the PDE system Eqs.~\eqref{eq:pde} and~\eqref{eq:tension} is solved simultaneously.

\section{\label{section:coarse} Asymptotic coarse-grained elastohydrodynamics}

In this section we describe the asymptotic coarse-graining formulation by integrating the moment balance system~\eqref{pde_1}-\eqref{pde_2}. The aim of this formulation is to bypass the complexity arising from the unknown contact forces~\eqref{eq:n}, not defined constitutively, thus requiring the Lagrange multiplier $\tau$ to ensure inextensibility~\eqref{eq:tension}. Integrating the balance of contact forces over the whole filament~\eqref{pde_1}, we get
\[
\mathbf{n}(L)-\mathbf{n}(0)+\int_0^L \mathbf{f}(s)\, \text{d}s=0,
\]
where external contact forces are given by $\mathbf{n}(L)=\mathbf{n}(0)=0$. The filament is conveniently divided in $N$ rod-like segments with $\Delta s = L/N$. In the asymptotic limit of a small, but finite, nonzero $\Delta s$, the filament can be coarse-grained via a semi-Riemann sum 
\begin{equation}
\sum_{i=1}^{N} \int_{(i-1) \Delta s}^{i \Delta s} \mathbf{f}(s) \, \text{d}s=\sum_{i=1}^{N} \mathbf{F}_{i}=0, 
\label{eq:discrete_Fbalance}
\end{equation}
so that, $\mathbf{F}_{i}$ represents the total contact force experienced by the $i$-th element. For a filament free from external torques, $\mathbf{m}(L)=\mathbf{m}(0)=0$, the total moment balance~\eqref{pde_2} simply reads 
\[
\sum_{i=1}^{N} \int_{(i-1) \Delta s}^{i \Delta s} \mathbf{x}_{s}(s) \times \mathbf{n}(s)\, \text{d}s =0. 
\]
After partial integration, and exploiting the force balance~\eqref{pde_1}, we find
\begin{equation}
\sum_{i=1}^{N} \int_{(i-1) \Delta s}^{i \Delta s} (\mathbf{x}(s)-\mathbf{x}_0) \times \mathbf{f} (s)\text{d}s = \sum_{i=1}^{N} \mathbf{M}_{i,\mathbf{x}_0} =0,
\label{eq:discrete_Mbalance}
\end{equation}
which is independent of $\mathbf{n}(s)$. Similarly, $\mathbf{M}_{i,\mathbf{x}_0}$ is the $i$-th moment about $\mathbf{x}_0=\mathbf{x}(0)$. As required, the total moment balance above is independent of the bending moment. Integration by parts of~\eqref{pde_2} for the $j$-th element instead introduces the effect of the  elastic bending moments via
\begin{equation}
 \sum_{i=j}^N \mathbf{M}_{i,\mathbf{x}_j}=\mathbf{m}_{j},
\label{eq:discrete_Me_balance}
\end{equation}
where $\mathbf{m}_{j} = \mathbf{m}((j-1)L/N) $ and $j = 2, \dots, N$. Here, it is convenient to write the moment $\mathbf{M}_{i,\mathbf{x}_j}$ relative to $\mathbf{x}_j$, whilst $\mathbf{m}_{j}$ is the bending moment contribution from the $j$-th element and, as previously, it is linearly related to the curvature $\mathbf{m}(s)= E_b \theta_s \mathbf{e}_z$. Distinct finite difference approximations maybe employed for $\theta_s$ \cite{Gadelha2010, delmotte_general_2015,Tornberg2003}. For simplicity, we use the backward difference formulae, 
\begin{equation}
\mathbf{m}_i=\kappa \, \alpha_i \mathbf{e}_z = \kappa \, (\theta_i - \theta_{i-1}) \mathbf{e}_z,
\label{eq:discrete_Melastic}
\end{equation}
where $\kappa=E_{b}/\Delta s$. The contact force $\mathbf{f}(s)$ in~\eqref{eq:discrete_Fbalance}-\eqref{eq:discrete_Me_balance} is given by the hydrodynamic coupling~\eqref{eq:fhydro}. 

We introduce now the geometry of deformation for centerline $\mathbf{x}(s,t)$ for the the coarse-grained elastohydrodynamic system. It is convenient to describe filament centerline in terms of the tangent angle $\theta$, see Fig.~\ref{schema_filament}, where $\mathbf{x}(s,t) = \mathbf{x}_{0} + \int_{0}^{L} (\cos\theta,\sin\theta)\text{d}s$, so that in the coarse-graining limit, we have
\begin{equation}
\mathbf{x}_i = \mathbf{x}_0 + \sum_{k=1}^{i-1} (\cos\theta_{k},\sin\theta_{k})\Delta s
\label{eq:discrete_x}
\end{equation}
for $i = 1, \dots, N$, thus $\mathbf{x}_i=\mathbf{x}((i-1)L/N)=(x_i,y_i)$, where $\theta_{k}$ is the angle between $\mathbf{e}_{x}$ and $\mathbf{e}_{k,\parallel}$ of the $k$-th element, $
\mathbf{e}_{k,\parallel}= ( \cos \theta_k, 
\sin \theta_k), \mathbf{e}_{k,\bot}=( -\sin \theta_k, \cos \theta_k)$, thus ensuring inextensibility intrinsically. Due to the curvature dependence in~\eqref{eq:discrete_Me_balance}, it is simpler to define the tangent angle in terms of the backward difference angle, $\alpha_i = \theta_{i}-\theta_{i-1}$, i.e. the angle between $\mathbf{e}_{i-1,\parallel}$ and $\mathbf{e}_{i,\parallel}$,
\begin{equation}
\theta_i = \sum_{k=1}^{i} \alpha_k,
\label{eq:discrete_alphathetha}
\end{equation}
by setting $\alpha_1 = \theta_1$. This reduces the filament centerline $\mathbf{x}(s,t)$ to only $N+2$ parameters $( x_0, y_0, \alpha_1, \dots , \alpha_N)$ (see \cite{alouges_self-propulsion_2013}). The total force balance~\eqref{eq:discrete_Fbalance} and torque balance~\eqref{eq:discrete_Mbalance}, together with $N-1$ equations for the internal moment balance~\eqref{eq:discrete_Me_balance}, further closes the elastohydrodynamic system with $N+2$ scalar equations.

The resistive force theory approximation~\eqref{eq:fhydro} allows for further analytical progress, as described in the seminal work by Gray and Hanckock \cite{GrayHancock55}, by expressing the anisotropic operator in terms of tangent angle. Thus $\mathbf{F}_{i}$ and $ \mathbf{M}_{i,\mathbf{x}_j}$ can be integrated analytically over the coarse-grained elements and expressed in terms of $(\dot{\mathbf{x}}_i,\dot{\theta}_i)$, where the overdots represent time derivatives. For simplicity, we assume linear interpolation of the shape function along the length $s'$ of each coarse-grained element.  Thus from ~\eqref{eq:discrete_x} the velocity of the centreline can be expressed as 
$$
\dot{\mathbf{x}}_{i}(s)= \mathbf{x}_i + (s-(i-1)\Delta s)\, \dot{\theta}_i  \mathbf{e}_{i,\bot}. 
$$
At the fixed frame of reference, the contact forces over the $i$-th coarse-element reads \cite{alouges2015can,GiraldiPomet16}
\begin{equation}
\mathbf{F}_i = \eta \Delta s \; \Lambda(\theta_i)^T \begin{pmatrix}  \dot{\mathbf{x}}_i \\ \Delta s \, \dot{\theta}_i \end{pmatrix},
\label{eq:Fi}
\end{equation}
where
$$
\Lambda(\theta) = \begin{pmatrix} -\cos^2 \theta-\gamma \sin^2 \theta & (\gamma-1)\cos \theta \sin \theta \\ (\gamma-1)\cos \theta \sin \theta & -\gamma \cos^2 \theta- \sin^2 \theta \\ \frac{1}{2} \sin \theta & -\frac{1}{2} \cos \theta \end{pmatrix}.
$$
Similarly, the contact moment at the $i$-th element relative to $\mathbf{x}_j$ takes the form 
\begin{equation}
\mathbf{M}_{i,\mathbf{x}_j} = \eta \Delta s \begin{pmatrix} 
\Delta s \\ x_i-x_j \\ y_i-y_j \end{pmatrix} ^T 
G(\theta_i)
\begin{pmatrix} \dot{x}_i \\ \dot{y}_i \\ \Delta s \, \dot{\theta}_i \end{pmatrix} \mathbf{e}_z 
\label{eq:Mi}
\end{equation}
with 
\begin{equation}
G(\theta) = \begin{pmatrix}
\scriptstyle - \frac{1}{2} \cos \theta 
& \scriptstyle \frac{1}{2} \sin \theta
& \scriptstyle -\frac{1}{3}
\\ \scriptstyle (1-\gamma) \cos \theta \sin \theta
& \scriptstyle - \cos^2 \theta - \gamma \sin^2 \theta
& \scriptstyle -\frac{1}{2} \cos \theta
\\ \scriptstyle \gamma \cos^2 \theta + \sin^2 \theta
& \scriptstyle (\gamma-1)\cos \theta\sin \theta
& \scriptstyle - \frac{1}{2} \sin \theta
\end{pmatrix},
\label{matrix_torques}
\end{equation}
where the above set of $3N$ variables $ \mathbf{X}_{3N}=(x_1,\dots,x_N, y_1,\dots, y_N, \theta_1,\dots,\theta_N)$ can be reduced to $N+2$ variables $ \mathbf{X}=(x_1,y_1,\theta_1,\dots,\alpha_N)$ via ${\mathbf{X}}_{3N}= {\mathbf{Q}} {\mathbf{X}}$, as described in detail in the Appendix. The coarse-grained elastohydrodynamics ~\eqref{eq:discrete_Fbalance}-\eqref{eq:Mi} reduces to a nondimensional system of ordinary differential equations 
\begin{equation} 
\mathrm{Sp}^4 \mathbf{A} \mathbf{Q} \dot{\mathbf{X}}=\mathbf{B},
\label{eq:ODE}
\end{equation}
where $\mathrm{Sp}$ is the sperm number as defined in~\eqref{eq:pde}, following the same recalling used for the classical system in the previous section. The general form of the matrices $\mathbf{A}$ and $ \mathbf{B}$ are also defined in the Appendix.

\begin{figure*}
\begin{center}
\includegraphics[width=\textwidth]{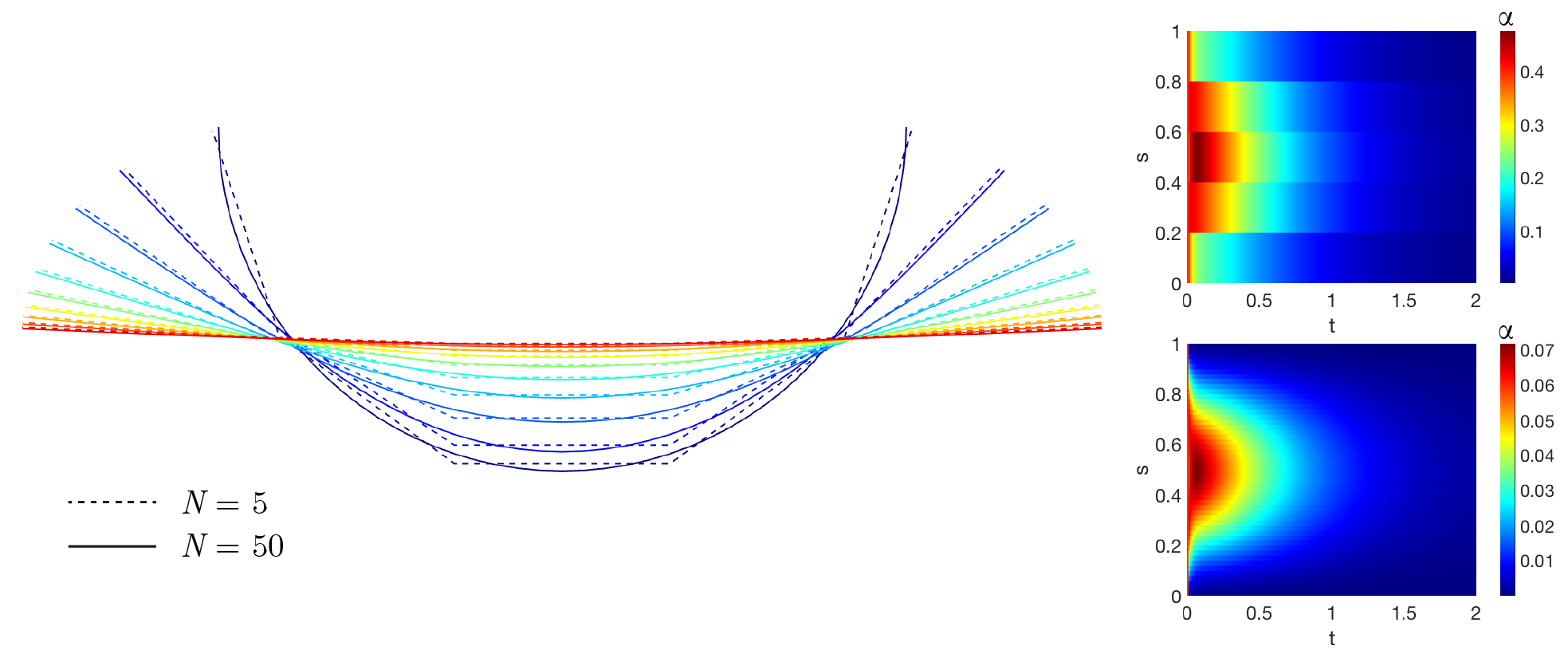}
\caption{Relaxation dynamics of a filament from a half-circle configuration with $5$ elements (dotted line) and $50$ elements (continuous line) and $Sp = 4$. The filaments are drawn on the left plot at $t=0$ (dark blue) and at regular time increment of $0.2$. The colormaps on the right show the spatiotemporal transient dynamics of the angle $\alpha_i = \theta_i - \theta_{i-1}$ between consecutive elements, i.e. the discrete curvature. The colormap above (below) corresponds to the coarse-grained filament with  $N=5$ ($N=50$) elements. Note that the values of $\alpha$ are smaller for the finer case,  as refinement induces a smoother spatiotemporal map, hence smaller angle difference between segments.}
\label{snaps_relax}
\end{center}
\end{figure*}

\section{Comparison between the classical and coarse-grained formulations \label{numerical_comp}}

The classical elastohydrodynamic system is solved using the numerical scheme used in \cite{gadelha_nonlinear_2010,Tornberg2003,montenegro-johnson_spermatozoa_2015}, briefly described here for comparison purpose. The system~\eqref{eq:pde}-\eqref{eq:tension} couples nonlinearly a fourth-order partial differential equations with a second-order boundary value problem for the unknown line tension, yielding severe constraints for the time-stepping size if all terms are treated explicitly \cite{gadelha_nonlinear_2010,Tornberg2003}. This is resolved by employing a second-order implicit-explicit method (IMEX) \cite{ascher_implicit-explicit_1995}, where only the higher-order terms are treated implicitly, and before any previous time level is available, the second-order IMEX is replaced by the first-order IMEX [ibid]. The arclength discretization is uniform with $N$ intervals, while second-order divided differences are used to approximate spatial derivatives, in which skew operators are applied at the boundaries \cite{gadelha_nonlinear_2010,Tornberg2003}. The timestep thus can be chosen to be the same order of magnitude as the grid spacing, yielding a first-order constraint for timestepping. Each iteration is made in two steps: first the boundary value problem for the Lagrange multiplier $\tau$, Eq.~\eqref{eq:tension}, is solved for a given filament configuration $\mathbf{x}$ at time $t_n$, from which Eq.~\eqref{eq:pde} can be timestepped to obtain new filament configuration $\mathbf{x}$ at $t_{n+1}$. Theoretical and empirical validation of this scheme is provided in Refs. \cite{gadelha_nonlinear_2010,Tornberg2003,montenegro-johnson_spermatozoa_2015}.

The coarse-grained elastohydrodynamic system~\eqref{eq:ODE} does not require evaluation of Langrange multipliers. The inextensibility is satisfied by model construction, while the asymptotic corse-graining allows for a straightforward semi-analytic relation between the filament kinematics and the elastohydrodynamic forces and torques. The ODE system~\eqref{eq:ODE} is straightforward to implementation using any solver or numerical scheme of choice. To illustrate this, we solved~\eqref{eq:ODE} using the built-in \texttt{ode15s} \textsc{Matlab} solver, which uses a variable-order, variable-step method based on the numerical differentiation formulas of orders 1 to 5 \cite{shampine1997matlab}. All computations for both the classical and coarse-grained formulation were conducted on an Intel Core i5-6500 processor 3.20 GHz, using \textsc{Matlab} software.

\begin{figure*}
\begin{center}
\includegraphics[width=\textwidth]{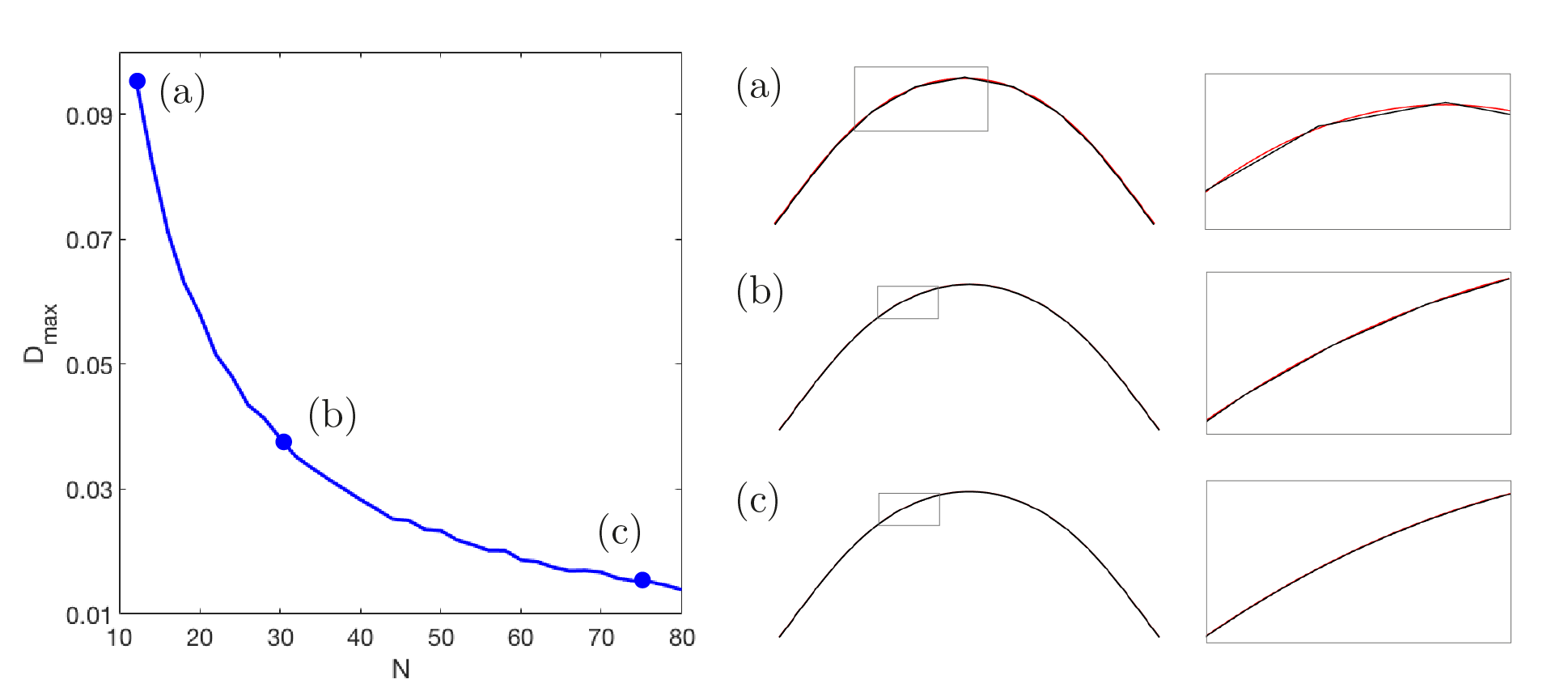}
\caption{Comparison between the classical and of the coarse-grained systems for increasing $N$. A fine discretisation for the classical solution is fixed for all cases. (a-c) show the aspect compare the classical (red) and the corse-grained (black) for $N = 10, 30, 75$, as indicated on the $D_{\text{max}}$ plot. The right-hand column exposes the detailed shape of small parts of the filaments, indicated by the small gray rectangles. The good accuracy of the corse-grained solution is observed even for very small number of segments $N$.}
\label{fig:Dist}
\end{center}
\end{figure*}

In what follows, we study the transient dynamics of a filament decaying from an initial configuration \cite{coy_counterbend_2017,Wiggins1998a,Wiggins2}, set to be a half-circle and a parabola. The filament thus unbends to its straight equilibrium configuration. Fig. \ref{snaps_relax} contrasts the coarse-grained filament configuration and difference angle $\alpha$ for $N=5$ and $N=50$. A remarkable agreement between the dynamics of a very coarse filament (with only 5 segments) and $N=50$ is observed. Indeed, the bulk-part elastohydordynamics is well captured by the the coarser system. This is despite the shape inaccuracies associated with high curvatures. The shape discrepancies are continuously reduced as the the filament approaches the equilibrium state. A higher number of segments smooths the elastohydrodynamic hyperdiffusion profile, thus acting as an effective spatial spline interpolation for each filament configuration in time (compare the angle plots in Fig. \ref{snaps_relax}). De facto, the coarse-grained system is able to capture the filament elastohydrodynamics with excellent accuracy even for very coarse filaments when compared with the classical system. This is agreement with Fig.~\ref{fig:Dist} which depicts the discrepancy between the classical, $\mathbf{x}_{\text{c}}$, and the coarse-grained, $\mathbf{x}_{\text{cg}}$, solutions via
\[
 D_{\text{max}} =  \max\limits_{s,t} | \mathbf{x}_{\text{c}} - \mathbf{x}_{\text{cg}}|,
\] so that $D_{\text{max}}=0$ if the agreement is exact \cite{Gadelha2010}. For $D_{\text{max}} \approx 0.1$ or less, the agreement is observed to be very good, as illustrated by the shapes in (a)-(c) for an increasing $N$ (Fig \ref{fig:Dist}). For $D_{\text{max}}<0.05$ the difference between the classical and coarse-grained solutions is almost undistinguishable, see for example the detailed insets in Fig.~\ref{fig:Dist}(b,c) on the right column. $D_{\text{max}} $ decays approximately with $1/N$ in Fig \ref{fig:Dist}, as expected from linear interpolation of curves. The dynamics is thus weakly influenced by the coarse-graining refinement level of the system. This feature may be exploited to reduce the dimencionality of the linear system while keeping a reasonable accuracy of the dynamics. By construction the asymptotic integrals along coarse-grained segments will tend to zero for infinitesimal $\Delta s$, as detailed in Eqs.~\eqref{eq:discrete_Fbalance} and~\eqref{eq:discrete_Mbalance}. This introduces a higher bound for $N$. For $N>80$, or equivalently for $\Delta s/ L < 1\%$, the system becomes numerically stiff and requires an excessive time-stepping refinement. 

We further compare the computational time of both formulations  in Table~\ref{table_results}. We focus on the numerically stiff regime of the classical  system, occurring at low sperm number $\mathrm{Sp}$, for effectively stiff filaments, and high curvatures. $N=70$ was used for all simulations of the coarse-grained model. The classical system however requires distinct spatiotemporal discretizations according to total length error associated to each parameter regime \cite{Tornberg2003}, chosen to give the smallest computing time. Table \ref{table_results} shows that the coarse-grained model has a maximum time duration of 3 seconds for $\mathrm{Sp}=2$. The computational time for the coarse-grained system increases as the $\mathrm{Sp}$ is reduced, although the accretion is marginal. On the other hand, the classical system suffers dramatically from numerical stiffness. For the lowest length-error tolerance imposed, $1\%$, the computational time increases by a factor of $74$ for the half-circle case when $\mathrm{Sp}$ is reduced. The time required for the the parabola is on the order of hours. The latter is exacerbated when length-error tolerance is reduced to $0.01\%$. In this case, even for $\mathrm{Sp}=4$, the computational times surpasses one hour to solve the parabola initial shape. De facto, this regime is known to be numerically challenging, as one approaches the limit of validity of the resistive-force theory. Elastic forces and torques are very large compared to the viscous dissipation, characterised by a snap-through, fast unbending of the filament towards the relaxation state, thus requiring very fine time-stepping to resolve this fast transient. Table \ref{table_results} demonstrates  how the coarse-grained approach outperforms the classical elastohydodynamic system.

\begin{table}
\begin{center}
\begin{tabular}{|c|c|c|c|}
\hline
\multicolumn{4}{|c|}{ \large $\mathrm{Sp}=4$} \\
\hline
Test & \footnotesize Coarse-grained & Tolerance & Classical\\ 
& \footnotesize system $N$=70 & \footnotesize length error & system \\
\hline
Half-circle & & 1\% & 1.3 \\
\multirow{2}{*}{\includegraphics[width=0.05\textwidth]{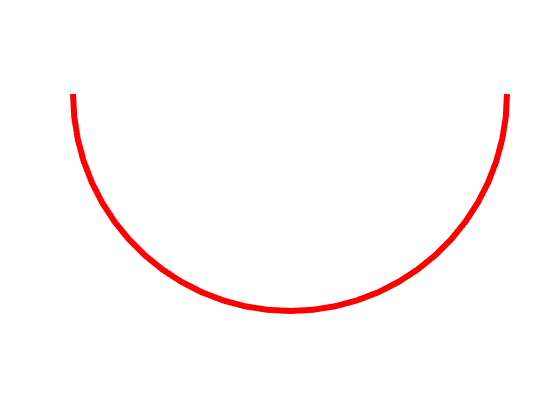}} & 2 & 0.1\% & 249 \\
 &  & 0.01\% & 3750 \\
\hline
Parabola & & 1\% & 90 \\
\multirow{2}{*}{\includegraphics[width=0.05\textwidth]{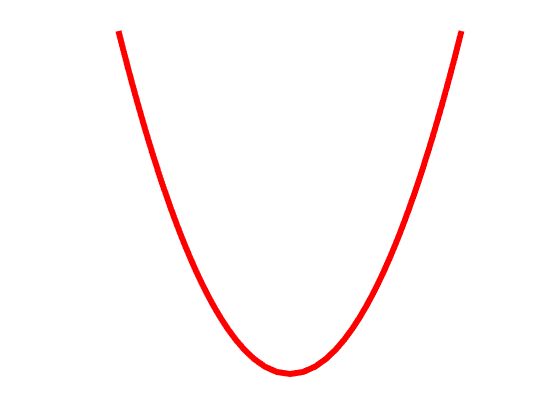}} & 1.5 & 0.1\% & 1820 \\
& & 0.01\% & $>1$h \\
\hline
\multicolumn{4}{|c|}{ \large $\mathrm{Sp}=2$} \\
\hline
Test & \footnotesize Coarse-grained & Tolerance & Classical\\ 
& \footnotesize system $N$=70 & \footnotesize length error & system \\
\hline
Half-circle & &1\% & 97 \\
\multirow{2}{*}{\includegraphics[width=0.05\textwidth]{half_circle.png}} & 3 & 0.1\% & 850 \\
& & 0.01\% & $>1$h \\
\hline
Parabola & & 1\% & $>1$h \\
\multirow{2}{*}{\includegraphics[width=0.05\textwidth]{parabola.png}} & 1.7 & 0.1\% & $>1$h \\
& & 0.01\% & $>1$h \\
\hline
\end{tabular}
\caption{Computing times in seconds for the two relaxation tests and two different sperm numbers. Tolerance of the length error only applies to the classical system. }
\end{center}
\label{table_results}
\end{table}

\section{Bio-applications \label{applications}}

In this section we apply the coarse-grained formulation for a variety of elastohydrodynamic systems and boundary conditions found in biology, emphasizing the simplicity and robustness of this approach. We focus on the filament buckling problem (Subsection \ref{section:buckling}), well known for its numerical stiffness, instability and challenges associated with boundary forces. We also study the magnetic actuation of swimming filament (Subsection \ref{sub:Mag}) and the dynamics of cross-linked filament-bundles and flagella (Subsection \ref{sub:cross}), including explicit elastic coupling among coarse-grained filaments.  Other interactions, via boundary forces/torques or their distribution along the filament, such as in gravitational and electromagnetic effects, as well as background flows, may be accounted effortlessly within this formulation.

\begin{figure*}[t!]
\begin{center}
\includegraphics[width=\textwidth]{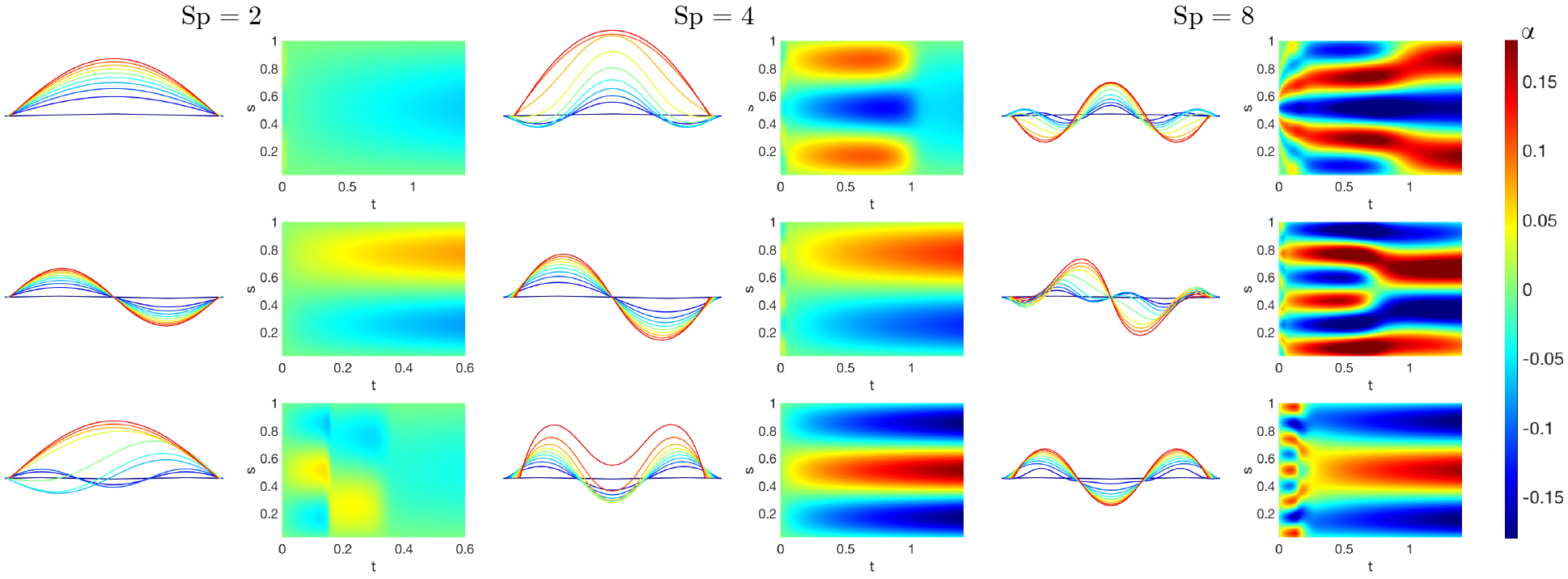}
\caption{Visualization of the buckling phenomenon for three different sperm numbers (Sp$=2$ on the left, Sp$=4$ on the middle and Sp$=8$ on the right). Here $N=36$ and $k=1.5$. The filament is displayed at regular intervals of time, coloured from blue (beginning) to red (end). Three different initial conditions lead to different outcomes. The colormap graphs show time on the x-axis and link number (discrete arclength) on the y-axis. The colours show the curvature (angles) over time, from blue for a highly negative curvature to red for a highly positive one. The filament quickly takes a waveform with fewer waves as time goes on. The higher the sperm number is, the longer it takes for the waves to vanish (note that the total time on the graphs is longer as the sperm number increases).}
\label{fig:buckling}
\end{center}
\end{figure*}

\subsection{Filament buckling instability \label{section:buckling}}

The coarse-grained system Eq.~\eqref{eq:ODE} is particularly suitable for non-trivial boundary constraints, such as in a fixed or moving boundary cases. In such situations, either the position (angle) or the force (torque) is imposed at the extremities. Here we consider an initially straight filament with the proximal end, $s=0$, pinned, so that the position is fixed but free from external torques. The distal boundary, $s=1$, moves with an imposed velocity towards the proximal end, although free from external torques. Post-transient dynamics, at the steady-state, leads to the celebrated Euler-elastica boundary value problem which admits exact solutions in terms of elliptic functions \cite{book:antman,Gadelha_elastica}. In this limit, contact forces balance exactly the the imposed load, and the shape is defined by the torque balance \cite{book:antman,Gadelha_elastica}. The transient dynamics of a filament buckling in a viscous fluid however depends to the distribution of both contact forces and toques that evolves in time. This requires the evaluation of unknown boundary forces at the proximal end, $s=0$, while the distal end, $s=1$, follows prescribed kinematics. We consider that the two endpoints are driven towards each other at a constant speed. This is a nontrivial task within the classical elastohydrodynamic formulation, as the usual separation between Eqs.~\eqref{eq:pde} and~\eqref{eq:tension} for the customary free force/torque condition is not possible. Instead, the unknown tension line at the boundary, required for the inextensibility constraint, is non-linearly coupled with the hyperdiffusive elastohydrodynamics \eqref{eq:pde}. This difficulty is augmented by the fact that the buckling instability is instigated by excessive compressive force distribution to a critical level in which the filament cannot uphold and  buckles. This occurs via a pitchfork bifurcation with equal chances to buckle in either direction, as $\theta \rightarrow -\theta$ is also a solution \cite{Gadelha_elastica}. The initial straight configuration thus requires an infinitesimal bias to trigger the unstable modes dynamically. 

The buckling phenomena is however straightforward within the coarse-grained framework. For this, we introduce unknown contact forces, respectively, for the proximal and distal ends $\mathbf{P}_0$ and $\mathbf{P}_N$, and associated moments in the coarse-grained system Eq.~\eqref{eq:ODE}, which reduces to 
\begin{equation}
\begin{array}{r l}
\displaystyle \sum_{i=1}^{N} \mathbf{F}_{i} + \mathbf{P}_0 +\mathbf{P}_N & = 0 \nonumber \\ 
\displaystyle \sum_{i=j}^N \mathbf{M}_{i,\mathbf{x}_j} + \mathbf{L}_{N,\mathbf{x}_j} & =\mathbf{m}_{j}\delta_{j},
\end{array}
\label{eq:Fbalance_buckling}
\end{equation}
for $j=1,\dots,N$ and $\delta_{j}$ is defined $\delta_{1}=0$ and $\delta_{j\neq1}=1$, where $\mathbf{L}_{N,\mathbf{x}_i}$ is the moment induced by $\mathbf{P}_N$ with respect to the point $\mathbf{x}_i$, $\mathbf{L}_{N,\mathbf{x}_i}=(\mathbf{x}_{N+1} - \mathbf{x}_i ) \times \mathbf{P}_N$. The unknown forces are supplemented by the kinematic constraints $\dot{\mathbf{x}}_{0} = (k/2,0)$ and $\dot{\mathbf{x}}_{N} = (-k/2,0)$, where $k$ is positive parameter. The detailed form of the linear system may be found in the Appendix, section \ref{appendix:buckling}.

Fig.~\ref{fig:buckling} depicts the shape evolution for the first three initially unstable modes at the onset of the instability and beyond, for $\mathrm{Sp}=2,4,8$ from Eqs.~\eqref{eq:Fbalance_buckling}. They capture the fast transient solutions for an effectively stiff filament, $\mathrm{Sp}=2$ (also the numerically stiff case), which rapidly collapses into the static, steady-state Euler-elastica solutions for the first two modes, top and middle plots in Fig.~\ref{fig:buckling} for $\mathrm{Sp}=2$. Complex mode competition is easily accessible. This is demonstrated by the third mode dynamics (bottom row). The coarse-grained system thus unveils the cascade of unstable modes towards the stable shape, see third-mode for $\mathrm{Sp}=8$ . For $\mathrm{Sp}=2$, these transitions  occur via fast distal-proximal travelling waves, with distinct wave duration and speed, as demonstrated by the spatiotemporal-$\alpha$ profiles. As $\mathrm{Sp}$ increases, more unstable modes are instigated, giving rise to a wide diversity of nonlinear phenomena and interactions among the participating modes, in particular mode-coupling competition, see for exemple $\mathrm{Sp}=4,8$ in Fig.~\ref{fig:buckling}. Investigation of mode stability at advanced, nonlinear stages is also possible using this formulation, for instance, by studying  the energy landscape and bifurcation diagrams. Despite the current gap in the literature, detailed nonlinear investigation of the buckling phenomenon in a viscous environment is outside the scope of the present paper and will be explored elsewhere.

\subsection{Magnetic swimmer \label{sub:Mag}}

\begin{figure*}[t!]
\begin{center}
\includegraphics[width=\textwidth]{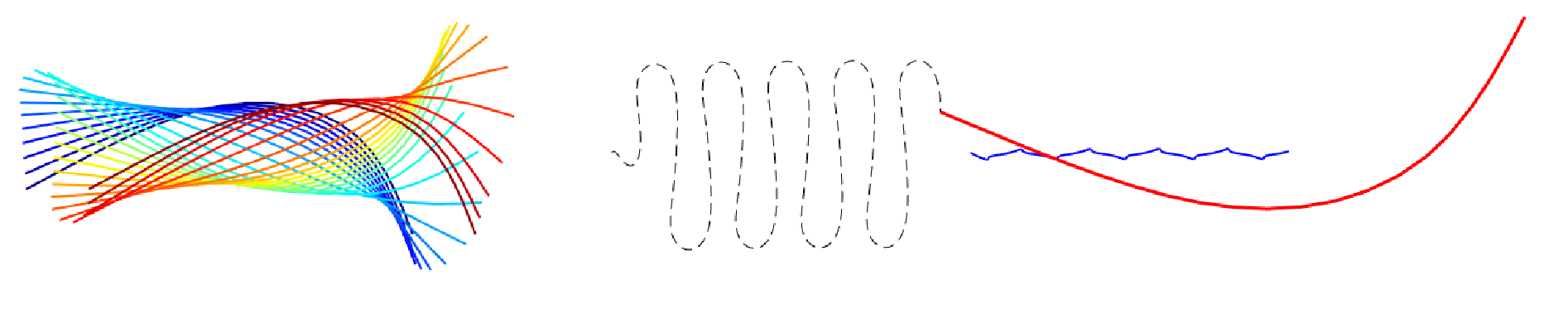}
\caption{Example of magnetic drive with a sinusoidal orthogonal magnetic field. One quarter of the length of the filament (i.e. the first five elements) is not magnetized, and the other part is constantly magnetized. Here $\mathrm{Sp}=4$, $N=20$, $M=1$, $\mathbf{H}(t)=\cos (t) \mathbf{e}_y /15$. On the left, the filament is represented at regular time intervals over a time period. On the right, the red line shows the position of the filament at the end of the simulation. The thin dotted and thick blue lines respectively show the trajectory of the non-magnetized end and the centroid of the filament.}
\label{fig:mag}
\end{center}
\end{figure*}

Following recent resurgence of interest on magnetically driven elastic fibres for the purpose of locomotion at micro or macro-scale  \cite{AlougesDeSimone15,dreyfus_microscopic_2005,gadelha2013optimal}, we solve the coarse-graining of a magnetic filament under the influence of an external magnetic field. In this section, we consider a magnetic filament with a homogeneous magnetic moment $\mathrm{N}$ along its arclength, directed towards the tangential direction, under the action of an uniform, time-varying sinusoidal oscillatory magnetic field $\mathbf{H}(t)$. The new terms arising from external torques are thus straightforward, as it only requires the addition of a distribution of the magnetic moments in Eq.~\eqref{eq:ODE}, $\mathbf{m}_i^m=\mathrm{N}_i \mathbf{e}_{i,\parallel} \times \mathbf{H}$, and reads
\begin{equation}
\begin{array}{r l}
\displaystyle \sum_{i=1}^{N} \mathbf{F}_{i} & =0 \,,\nonumber \\ 
\displaystyle \sum_{i=j}^N (\mathbf{M}_{i,\mathbf{x}_j} +\mathbf{m}_i^m) & =\mathbf{m}_{j}\delta_{j},\,\, \textrm{for j = 1 $\dots$ N}.
\end{array}
\label{eq:magnetic_swimmer}
\end{equation}
Fig.~\ref{fig:mag} shows an example of a partially magnetised swimmer moving according to the applied sinusoidal magnetic field, with $N=20$ and $\mathrm{Sp}=4$, starting from a straight configuration for approximately 5 cycles. The coarse-grained system is numerically cheap, as it has a reduced number of mesh points. Thus it allows for optimisation studies involving the continuous evaluation of objective function across a large parameter space. Previous studies demonstrated that the classical system leads to very expensive numerical simulations \cite{gadelha2013optimal,montenegro-johnson_spermatozoa_2015}, making any parameter search very challenging. This open new possibilities for investigations within control theory, as well as optimal control \cite{GiraldiPomet16, GiraldiMoreau17} by using this approach.

\subsection{Cross-linked filament bundles and flagella \label{sub:cross}}

In this section we focus on biological systems involving time-dependent load distributions. This could arise, for example, via mechano-sensory coupling in biological structures and biochemical landscapes.  Flagella and cilia found in eukaryotes are perfect exemplars of the latter \cite{gaffney}. They are composed by a geometrical arrangement of semi-flexible filaments interconnected by elastic linking proteins, called axoneme \cite{satir1974structural}. Its generic form is composed by $9+2$ microtubule doublets surrounding a central pair \cite{gadelha_counterbend_2013,lindemann2005counterbend,pelle2009mechanical}, observed in both motile and non-motile form. Flagella is a challenging mathematical system. It couples nanometric scales from the molecular motor biochemical activation with microscopic properties of the elastic structure, as observed for the purpose of spermatozoa transport \cite{gaffney,ishimoto_human_2018}. A geometrical abstraction of this system based on the sliding filament mechanism was first proposed by Brokaw  \cite{brokaw_flagellar_1972}. In the static case, for steady-state deformations, flagella is prone to the so-called \textit{counterbend phenomenon} \cite{lindemann2005counterbend,gadelha_counterbend_2013}.  This occurs when distant parts of a passive flagellum (in absence of motor activity) bend  in opposition to an imposed curvature elsewhere along the flagellum, for example, using the tip of a micropipete \cite{Lindemann2005,pelle2009mechanical,bayly_counterbend}.  Theoretical models encoding the mean cross-linked filament-bundle mechanics  where able to recover the counterbend phenomenon \cite{gadelha_counterbend_2013,Gadelha_elastica}, from which material parameters could be measured directly from the resulting counter-curvature. The dynamics of passive flagellar bundles have been investigated using linear theory \cite{gadelha_counterbend_2013}, and prediction of counter-travelling waves instigated by the non-local cross-linking moments reported. To date, a geometrically exact investigation is still lacking in the literature.

\begin{figure*}[t!]
\begin{center}
\includegraphics[width=\textwidth]{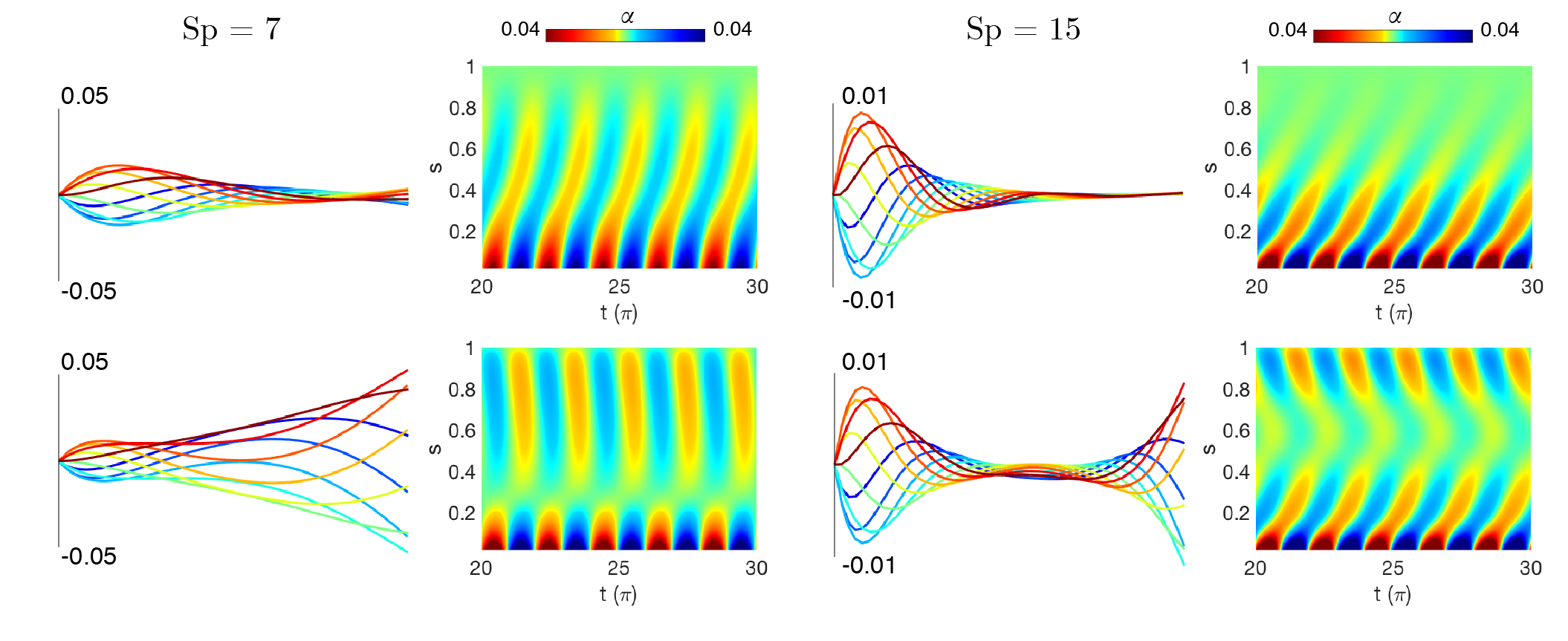}
\caption{Simulation of the counterbend phenomenon for two different sperm numbers (parameters are chosen in order to match those used in \cite{coy_counterbend_2017}: $\kappa^s/\kappa=0.06$, $a=0.4362\, \mathrm{rad}$; and $N=50$). The top row shows the behaviour of an actuated filament when no sliding resistance is added. The right column shows the same oscillation applied to the bundle model. The color plots show the curvature (angles $\alpha$) with respect to the time in $x$ and the arclength in $y$. The travelling curvature wave generated by the actuation is visible at the bottom of all of the color plots. For the counterbend case, a second travelling wave appears at the free end of the filament in the bundle case (bottom row).}
\label{fig:bundle1}
\end{center}
\end{figure*}

We consider the geometrically exact cross-linked filament bundle system for a passive bundle, that is a flagellum without molecular motor actuation, using the coarse-grained formulation.  The sliding filament model \cite{brokaw_flagellar_1972,coy_counterbend_2017} is particularly cumbersome within  the classical elastohydrodynamic framework \cite{gadelha_nonlinear_2010}.  The boundary conditions are non-local due to the accumulative dependence of sliding moments along the bundle, and generally unknown during the dynamics. This is becomes even more challenging when the bundle is driven via the molecular motor activity \cite{oriola_nonlinear_2017,coy_counterbend_2017}, which may explain the reason why numerical investigation of geometrically exact bundle systems are still lacking, for both active and passive cases \cite{oriola_nonlinear_2017}. The coarse-grained formulation brakes the contribution of the sliding filament moments for each segments simply as
\begin{equation}
\mathbf{m}_{i}^{s}=\kappa^s \sum_{j=i}^N ( \theta_i - \theta_1 ),
\end{equation}
where $\kappa^s$ is an effective resistance to sliding between the sliding filaments \cite{coy_counterbend_2017}. The balance of forces and moments then reads
\begin{equation}
\label{eq:Fbalance_bundle1}
\begin{array}{r l}
\displaystyle \sum_{i=1}^{N} \mathbf{F}_{i} & = 0\,, \nonumber \\ 
\displaystyle \sum_{i=j}^N \mathbf{M}_{i,\mathbf{x}_j} & =\mathbf{m}_{j}\delta_{j} +\mathbf{m}_j^s\,,\textrm{{for j = 1 $\dots$ N}},
\end{array}
\end{equation}
and  describes an effective sliding filament bundle free from forces/torques at endpoints. We consider instead  that the bundle is fixed and angularly actuated at the proximal end. Thus the first three equations in Eq.~\eqref{eq:Fbalance_bundle1} are replaced with the kinematic conditions $\dot{\mathbf{x}}_1=0$ and $\dot{\theta}=a \cos{t}$ for an angular amplitude $a$. Numerical solutions of the the coarse-graining system for a filament bundle angularly actuated at $s=0$ with amplitude $a=0.4362\, \mathrm{rad}$ and $\kappa^s/\kappa=0.06$ is shown in Fig. \ref{fig:bundle1}. They confirm  analytical prediction of counter-wave phenomenon from linear theory reported in (\cite{coy_counterbend_2017}, Fig.~2), where waves are instigated non-locally, and travel in opposition to the imposed angular oscillation. The wavespeed and amplitudes involved depend on the cross-linking elastohydrodynamic parameters and the sperm number, compare $\mathrm{Sp}=7$ and $\mathrm{Sp}=15$ in Fig. \ref{fig:bundle1}. It is worth notting that coarse-graining system in Eq.~\eqref{eq:Fbalance_bundle1} allows for straightforward generalisation to include different motor-control hypothesis - central for the current flagella and cilia self-organisation current debate \cite{coy_counterbend_2017,sartori_dynamic_2016,oriola_nonlinear_2017}.

\begin{figure*}
\begin{center}
\includegraphics[width=\textwidth]{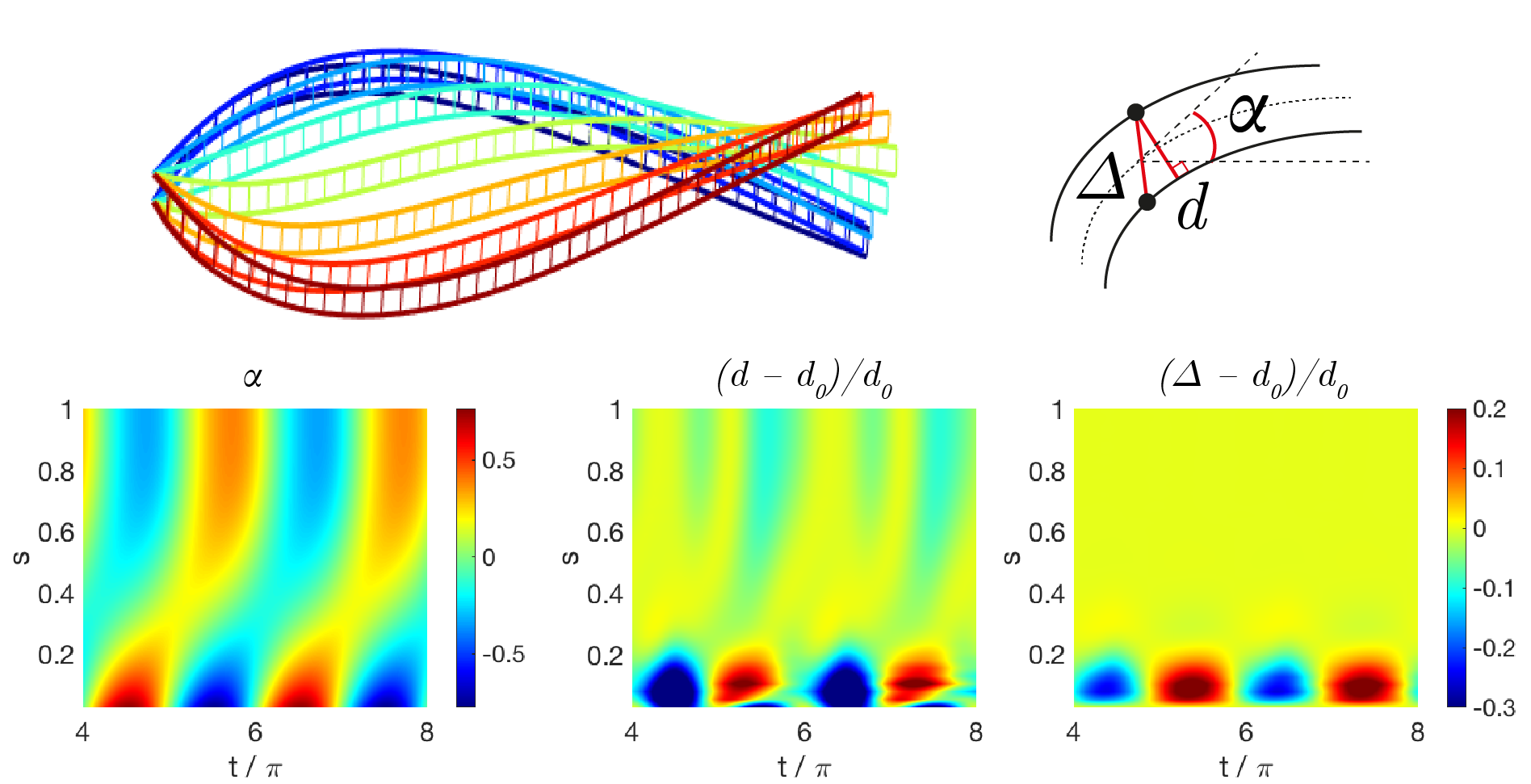}
\caption{Coupling between two filaments obtained with the coarse-grained approoach. Here $\mathrm{Sp}=4$, $a=0.88\, \mathrm{rad}$, $K/\kappa = 1/25$. In addition to the actuated filament represented at different times over half a time period (on the top), the three colormaps show three parameters with respect to time and arclength: the angle $\alpha$ of the centerline (left), the distance $d$ between the two filaments (middle) and the distance $\Delta$ between two facing nodes, normalised by their resting length $d_0$.
A travelling wave of curvature generated by the actuation of the top filament is visible in graph at the bottom left. The graph at the bottom in the middle shows the distance between the two filaments. Moreover, the graph at the bottom right captures the sliding distance between the two filaments.}
\label{fig:bundle2}
\end{center}
\end{figure*}

Finally, we consider the dynamics of two individual filaments embedded in a viscous fluid and coupled elastically via Hokean elastic springs, see Fig.~\ref{fig:bundle2}. The system thus involves the geometrically non-linear elastohydrodynamics of two interacting elastic fibres. Once again, the classical elastohydrodynamic formulation is ill-posed. The discrete distribution of elastic springs introduces  unknown point forces via Eqs.~\eqref{eq:pde} and~\eqref{eq:tension}  for each filament. We consider that the two-filament assembly is angularly actuated at one end, see Fig.~\ref{fig:bundle2}. We assume that the connecting elastic springs have an effective spring constant $K$ connecting oposite nodes between the two filaments, placed at each coarse-grained segment junction for simplicity, see  Fig.~\ref{fig:bundle2}. The equilibrium bundle diameter is $d_0$ at rest. The elastic force $\mathbf{F}^{\mathrm{int}}_{i}$ exerted by the $i$-th spring on the filament $S$ (top filament) reads
$$ \mathbf{F}^{\mathrm{int}}_{i} = K \, \left( 1 - \frac{d_0}{\| \mathbf{x}_i - \mathbf{x}'_i \|}\right) \, (\mathbf{x}_i - \mathbf{x}'_i),$$
where primes refer to the second filament. The coarse-grained formulation for $S$ and $S'$ is thus augmented by the elastic reactions from each connecting spring, and their associated moments along each filament. Hence the coupled system for the filaments $S$ and $S'$ reads, respectively,
\begin{equation}
\begin{array}{r l}
\displaystyle \sum_{i=1}^{N} (\mathbf{F}_{i} + \mathbf{F}^{\mathrm{int}}_{i}) & = 0 \nonumber \\ 
\displaystyle \sum_{i=j}^N (\mathbf{M}_{i,\mathbf{x}_j} + \mathbf{F}^{\mathrm{int}}_{i} \times (\mathbf{x}_j-\mathbf{x}_i) ) & = \mathbf{m}_{j} \delta_{j}, \\
\displaystyle \sum_{i=1}^{N} (\mathbf{F}'_{i} - \mathbf{F}^{\mathrm{int}}_{i}) & = 0 \nonumber \\ 
\displaystyle \sum_{i=j}^N (\mathbf{M}'_{i,\mathbf{x}_j} - \mathbf{F}^{\mathrm{int}}_{i} \times (\mathbf{x}'_j-\mathbf{x}'_i) ) & = \mathbf{m}'_{j} \delta_{j}.
\end{array}
\label{eq:Fbalance_bundle}
\end{equation}
The proximal end of both constituent filaments is fixed, but the angle at $s=0$ of the filament $S$ (top filament) is actuated via $\theta_1 (t) = a \sin t$.   The filament $S'$ (bottom filament) is free from external actuation, thus its movement solely arises via the elastic coupling between them. A detailed description of the resulting two-filament system is provided in the appendix, section \ref{appendix:bundle}.

Fig.~\ref{fig:bundle2} shows numerical simulations for time evolution of the two-filament assembly,  demonstrating the effectiveness of the connecting springs while transmitting bending moment from the top filament to the bottom one. A synchronous traveling wave of curvature is observed, see for exemple  the tangent angle $\alpha$ of the centreline of the filament-pair in Fig.~\ref{fig:bundle2}. The axial diameter $d$ however evolves asymmetrically (middle plot in Fig.~\ref{fig:bundle2}). The angular actuation of the top filament modify the diametral distance between the filaments near the base, in an oscillatory motion, from where axial waves are propagated down the structure. Axial extensional waves (light yellow regions) propagate more easily than  compressional waves in the axial direction (light green regions). The resulting sliding displacement $\Delta$ between the filament-pair is also depicted in the right graph in Fig.~\ref{fig:bundle2}. Similarly to the radial distance, the relative sliding motion is concentrated towards the basal end, however, it is not propagated along the filament-pair. This is despite the fact that both filament are inextensible, and tangential motion is easily propagated. Conversion of  curvature into relative sliding motion between the filaments is not observed nor the counterbend phenomenon observed in Fig.~\ref{fig:bundle1}, see the light yellow region for $\Delta \approx 0$ in Fig.~\ref{fig:bundle2}. This suggests that simple elastic connectors between filaments are not effective while transmitting sliding moments. Instead, axial distortions are prevailed and propagated along the filament-pair assembly. The connecting springs contribute to forces along its direction, but mostly on the radial direction, perpendicular to centreline. The elastic springs are hinged at each connecting node, thus the constituent filaments are free from bending moments arising from the interfilament sliding (in contrast with Fig.~\ref{fig:bundle1}). This supports the so-called geometric clutch mechanism proposed by Lindemann \cite{lindemann1994geometric}, where axial displacements are central for  flagellar mechanics. These results further show that the sliding filament mechanism present in flagellar systems \cite{brokaw_flagellar_1972,gadelha_counterbend_2013,lindemann2005counterbend} is far more complex than this simplistic two-filament cross-linked assembly \cite{Camalet,sartori_dynamic_2016,oriola2014subharmonic,coy_counterbend_2017,brokaw_flagellar_1972}.

\section{Conclusions}

This paper studies inertialess fluid-structure interaction of inextensible filaments commonly found in biological systems. The nonlinear coupling between the geometry of deformation and the physical effects invariably result on intricate governing equations that negotiate elastohydrodynamical interactions with non-holonomic constraints, as a direct consequence of the filament inextensibility. As a result, the classical elastohydrodynamical  formulation is prone to numerical instabilities, requires penalization methods and high-order spatiotemporal propagators. Here, we exploit the momentum balance in the asymptotic limit of small rod-like elements, from which the system can be integrated semi-analytically. This bypasses the bottleneck associated with the inextensibilty constraint, and does not require the use of Langrange multipliers to solve the system. We further show the equivalence between the two formalisms, as well as a direct comparison between the their numerical performances. The coarse graining formalism was shown to outperform the classical approach, in particular for numerically stiff regimes where the classical system performs poorly. The coarse-graining  structure also allowed faster computations, over than a hundred times faster than previous implementations. The coarse-graining approach is simple and intuitive to implement, and generalisations for complex interaction of multiple rods, active filaments, flagella, Brownian polymer dynamics and non-local hydrodynamics are straightforward. 

We employed the coarse-grained formulation to study distinct biologically inspired systems: the buckling instability of bio-filaments, the magnetic actuation of a microswimmer and the dynamics of cross-linked filament bundles and flagella. With the exception of the magnetic swimmer case, the results obtained here for the other systems are new in the literature. For the buckling problem, travelling waves are generated and propagated with different speeds, depending on the elastohydrodynamic properties of the filament and its interaction with other competing modes, Fig.~\ref{fig:buckling}; thus relevant to biological systems in which buckling is a naturally occurring phenomenon \cite{Bourdieu95}. The coarse-graining approach successfully captured the counterbend phenomenon in cross-linked filament-bundles \cite{coy_counterbend_2017,gadelha_counterbend_2013}, Fig.~\ref{fig:bundle1}, including geometrical nonlinearities. Finally,  motivated by mathematical abstractions of flagellar systems \cite{Camalet,sartori_dynamic_2016}, we solved the dynamics of interactions between two individual filaments interconnected with elastic springs. Numerical simulations indicated that the sliding between the filaments is not instigated by changes in curvature, as assumed by the sliding-filament mechanism \cite{brokaw_flagellar_1972}. Instead, axial distortions are  propagated along the two-filament assembly, in agreement with the geometric clutch hypothesis \cite{lindemann1994geometric}. These modes of deformation are central for the molecular-motor control debate in flagellar systems \cite{sartori_dynamic_2016,coy_counterbend_2017}.

The results presented here offer new possibilities for theoretical investigations, for instance, on the elastohydrodynamic self-organisation of fibres, many interacting filaments, cytoskeleton modelling, manoeuvrability of micro-magnetic robots \cite{GiraldiMoreau17,GiraldiPomet16}, as well as optimal strategy of deformation for micro-locomotion \cite{WiezelGiraldi18}. Only basic knowledge of systems of linear equations is required and implementation achieved with any solver of choice. We hope that the simplicity of the formalism, the numerical robustness and easy-to-implement generalisations will appeal to the biology, soft-matter and interdisciplinary community at large.

%
%
%
%
%
%
\section*{Acknowledgments}
The authors were partially funded by the CNRS through the project D\'{e}fi InFIniti-AAP 2017.
The authors also thank J.-B. Pomet for fruitful discussion.

\section{Appendix}
\subsection{Parametrization in the coarse-graining model \label{appendix_transofmration_matrices}}

In the asymptotic description, the filament can be described with two different sets of parameters (see figure \ref{schema_filament}) : 
\begin{itemize}
\item the $N+2$ parameters $\mathbf{X}=(x_1,y_1,\theta_1,\alpha_2,\dots,\alpha_N)$.
\item the $3N$ parameters $\mathbf{X}_{3N}=(x_1,\dots,x_N,y_1,\dots,y_N,\theta_1,\dots,\theta_N)$,
\end{itemize}
The second set uses $3N$ parameters where $N+2$ are sufficient. However, it makes the computations easier to read.
Going from $\dot{\mathbf{X}}$ to $\dot{\mathbf{X}}_{3N}$ can be done via the following transformation matrices :
$$ \dot{\mathbf{X}}=\tilde{\mathbf{P}} \dot{\mathbf{X}}_{3N} \quad \mathrm{and} \quad \dot{\mathbf{X}}_{3N}= \tilde{\mathbf{Q}} \dot{\mathbf{X}}
$$
with
$$
\tilde{\mathbf{P}}= \left (
\begin{array}{c|c|c}
\begin{matrix} 1 & 0 & \dots & & 0 \\ 0 & & \dots & & 0 \end{matrix} &
\begin{matrix} 0 & & \dots & & 0 \\ 1 & 0 & \dots & & 0 \end{matrix} &
0_{2,N} \\ \hline
0_{N} & 0_N &
\begin{matrix} 1 & 0 & \dots & \dots & 0 \\
       -1 & 1 & 0 & \dots & 0 \\
       0 & -1 & 1 & \ddots & \vdots \\
       \vdots & \ddots & \ddots & \ddots & \vdots \\
       0 & \dots & 0 & -1 & 1 
\end{matrix}
\end{array}
\right )
$$
and
\begin{equation}
\tilde{\mathbf{Q}}=\left (
\begin{array}{c|c}
\begin{matrix} 1 & 0 \\ \vdots & \vdots \\ 1 & 0 \end{matrix} & \tilde{Q}_1 \\ \hline 
\begin{matrix} 0 & 1 \\ \vdots & \vdots \\ 0 & 1 \end{matrix} & \tilde{Q}_2 \\ \hline
0_{N,2} & \begin{matrix} 1 & 0 & \dots & 0 
           \\ 1 & 1 & \ddots & \vdots 
           \\ \vdots & & \ddots & 0 
           \\ 1 & \dots & & 1 
\end{matrix}
\end{array}
\right) ,
\label{matrix_q}
\end{equation}
where $\tilde{Q}_1$ and $\tilde{Q}_2$ are $N \times N$ matrices whose elements are given by the general formula
\[
\begin{array}{l}
q_{1}^{i,j}=-\Delta s \sum_{k=j}^{i-1} \sin \left( \sum_{m=1}^{k} \alpha_m \right ) \\
q_{2}^{i,j}=\Delta s \sum_{k=j}^{i-1} \cos \left( \sum_{m=1}^{k} \alpha_m \right ),
\end{array}
\]
with $q_{1}^{i,j}=q_{2}^{i,j}=0$ if $i \geq j$. The tildes recall that the nondimensionalization has not yet been performed.

\subsection{Matricial form of the ODE system \label{appendix_matricial_form}}
Using the explicit expressions of the different contributions~\eqref{eq:discrete_Melastic},~\eqref{eq:Fi} and~\eqref{eq:Mi}, and after nondimensionalizing, we can rewrite the system~\eqref{eq:discrete_Fbalance}-\eqref{eq:discrete_Me_balance} under matricial form:
\begin{equation}
\mathrm{Sp}^4 \mathbf{A} \mathbf{Q} \dot{\mathbf{X}}=\mathbf{B}.
\label{edo_mat_nond}
\end{equation}
where the terms are defined as follows:
\begin{itemize}
\item The matrix $\mathbf{A}$ is a $(N+2) \times 3N$ matrix whose coefficients are given, for all $i$ in $\{ 1, \dots N \}$ and $j$ in $\{i,\dots, N\}$, by
\[
\begin{array}{l}
a_{1,i}= -\cos^2 \theta_i-\gamma \sin^2 \theta_i; \\ 
a_{2,i}=(\gamma-1)\cos \theta_i \sin \theta_i;\\
a_{1,N+i}=(\gamma-1)\cos \theta_i \sin \theta_i; \vspace{2mm} \\ 

a_{2,N+i}=-\gamma \cos^2 \theta_i- \sin^2 \theta_i;\\
a_{1,2N+i}=\frac{1}{2} \sin \theta_i; \\ 
a_{2,2N+i}=-\frac{1}{2} \cos \theta_i; \vspace{2mm} \\

a_{i+2,j}=v(\mathbf{x}_i,\mathbf{x}_j) \, M_1(\theta_j); \\ 
a_{i+2,N+j}=v(\mathbf{x}_i,\mathbf{x}_j) \, M_2(\theta_j); \\ 
a_{i+2,2N+j}=v(\mathbf{x}_i,\mathbf{x}_j) \, M_3(\theta_j);
\end{array}
\]
where $$ v(\mathbf{x}_i,\mathbf{x}_j)=\begin{pmatrix} 1 & \frac{x_j-x_i}{\Delta s} & \frac{y_j - y_i}{\Delta s} \end{pmatrix} $$ and $M_1, M_2, M_3$ are the columns of the matrix~\eqref{matrix_torques}. If $j<i$, then $a_{i+2,j}=a_{i+2,N+j}=a_{i+2,2N+j}=0$. 
\item $\mathbf{Q}$ is the nondimensionalized version of the transformation matrix~\eqref{matrix_q}. It is defined by replacing $\tilde{Q}_1$ and $\tilde{Q}_2$ with $Q_1=\tilde{Q}_1 / \Delta s$ and $Q_2 = \tilde{Q}_2 / \Delta s$ in the expression of $\tilde{\mathbf{Q}}$.
\item $\mathbf{B}$ is a column vector of size $N+2$, given by
$$
\mathbf{B}=\begin{pmatrix} 0 & 0 & 0 & \alpha_2 & \dots & \alpha_N \end{pmatrix}^T.
$$
\end{itemize}

\subsection{Buckling instability system \label{appendix:buckling}}

The resolution of the buckling problem requires to introduce two unknown contact forces $\mathbf{P_0}$ and $\mathbf{P}_N$ (see section \ref{section:buckling}). It yields four more scalar unknowns: $P_{0x}$, $P_{0y}$, $P_{Nx}$, $P_{Ny}$. We add four equations to the system~\eqref{eq:ODE} by embedding the buckling kinematic constraints $\dot{\mathbf{x}}_{0} = (k/2,0)$ and $\dot{\mathbf{x}}_N = (-k/2,0)$, where $k$ is positive. We get a new system of $(N+6)$ scalar equations that can be expressed in matricial form:
\begin{equation}
\mathbf{A}_b \dot{\mathbf{X}}_b = \mathbf{B}_b,
\label{edo_mat_buckling}
\end{equation}
where
\begin{gather}
\mathbf{X}_b = \begin{pmatrix} \mathbf{X} \\ \hline P_{0x} \\ P_{0y} \\ P_{Nx} \\ P_{Ny} \end{pmatrix}, \; \mathbf{B}_b= \begin{pmatrix} \mathbf{B} \\ \hline -k/2 \\ 0 \\ k/2 \\ 0 \end{pmatrix}, \; \mathbf{A}_b= \left ( \begin{array}{c|c} \mathrm{Sp}^4 \mathbf{A} \mathbf{Q} & \mathbf{a}^T \\ \hline \mathbf{a} & \mathbf{0} \end{array} \right ),
\\ 
\mathbf{a} = \begin{pmatrix} 1 & 0 & 0 & \dots & \dots & 0 
	\\ 0 & 1 & 0 & \dots & \dots & 0 
  \\ 1 & 0 & - \sum_{k=1}^N \sin \theta_k & -\sum_{k=2}^N \sin \theta_k & \dots & - \sin \theta_N
  \\ 0 & 1 & \sum_{k=1}^N \cos \theta_k & \sum_{k=2}^N \cos \theta_k & \dots & cos \theta_N \end{pmatrix},
\end{gather}

and matrices $\mathbf{A}$, $\mathbf{B}$ and $\mathbf{Q}$ are defined as in Appendix \ref{appendix_matricial_form}. 

\subsection{Magnetic swimmer}
The matricial system describing a magnetically driven filament with the coarse-graining approach reads
\begin{equation}
\mathrm{Sp}^4 \mathbf{A} \mathbf{Q} \dot{\mathbf{X}}=\mathbf{B} + \frac{1}{\kappa}\mathbf{C}^m.
\end{equation}
It is simply obtained by adding to the system~\eqref{eq:ODE}, the magnetic effect vector $\mathbf{C}^m = \begin{pmatrix} c_1 & \dots & c_{N+2} \end{pmatrix}^T$, with $c^m_1=c^m_2=0$ and $\forall i \in \{ 1, \dots, N \}$,
\[
c^m_{i+2} = \sum_{k=i}^{N} N_k (H_y (t) \cos \theta_k - H_x (t) \theta_k),
\]
$H_x (t)$ and $H_y (t)$ being the components of the magnetic field along the $x$- and $y$-axis.

\subsection{Cross-linked filament bundle \label{appendix:bundle}}
The system~\eqref{eq:Fbalance_bundle} describing a filament bundle with sliding resistance takes the following matricial form:
\begin{equation}
\mathrm{Sp}^4 \mathbf{A} \mathbf{Q} \dot{\mathbf{X}}=\mathbf{B} + \frac{\kappa^s}{\kappa} \mathbf{C}^s,
\end{equation}
with $\mathbf{C}^s = \begin{pmatrix} c_1 & \dots & c_{N+2} \end{pmatrix}^T$, with $c^s_1=c^s_2=0$ and $\forall i \in \{ 1, \dots, N \}$,
\[
c^s_{i+2} = \theta_i-\theta_1.
\]

In the case of two interacting filaments, $S$ and $S'$, the new coupled dimensionless system of $(2N+4)$ equations reads
\begin{equation}
\mathrm{Sp}^4  \! \left ( \! \!  \begin{array}{c|c}
(\mathbf{AQ})_{S} & 0 \\
\hline
0 & (\mathbf{AQ})_{S'} \end{array} \! \! \right ) \! \! 
\left ( \begin{array}{l} \mathbf{X}_{S} \\ \mathbf{X}_{S'} \end{array} \! \! \right ) \! = \! 
\left ( \begin{array}{l} \mathbf{B}_{S} \\ \mathbf{B}_{S'} \end{array} \! \! \right) \! + \! 
\left ( \! \begin{array}{l} \mathbf{C} \\ \mathbf{C}' \end{array} \! \right).
\label{eq:matricial-bundle}
\end{equation}
In the above equation, $\mathbf{A}$, $\mathbf{B}$, $\mathbf{Q}$ and $\mathbf{X}$ are defined as previously, referring to $S$ or $S'$ depending on the indix. The interaction vectors $\mathbf{C}$ and $\mathbf{C}'$ are defined as follows:
\begin{equation}
\begin{array}{l l}
& \begin{pmatrix} c_1 \\ c_2 \end{pmatrix} = \frac{1}{\kappa} \sum_{j=1}^{N} \mathbf{F}^{\mathrm{int}}_{j}; \\
& \begin{pmatrix} c'_1 \\ c'_2 \end{pmatrix} = - \frac{1}{\kappa} \sum_{j=1}^{N} \mathbf{F}^{\mathrm{int}}_{j}; \\
\forall i \in \{1, \dots N \}, & c_{i+2}= \frac{1}{\kappa} \sum_{j=i}^N \mathbf{F}_{j}^{\mathrm{int}} \times (\mathbf{x}_j - \mathbf{x}_i ), \\
& c'_{i+2}= - \frac{1}{\kappa} \sum_{j=i}^N \mathbf{F}_{j}^{\mathrm{int}} \times (\mathbf{x}'_j - \mathbf{x}'_i ).
\end{array}
\end{equation}

The above system describes a bundle with free endpoints. However, in the case studied in section \ref{eq:matricial-bundle}, both filaments have a fixed proximal end, and the filament $S$ is actuated at its proximal end (prescribed angle $\theta_1$). We embed these boundary conditions in the system by replacing its first three lines by the constraint equations $\dot{x}_1 = 0$, $\dot{y}_1 = 0$, $\dot{\theta}_1 = a \cos t$, and its $N+3$-th and $N+4$-th equations (i.e., the first two equations for the second filament) by the constraint equations $\dot{x}'_1 = 0$ and $\dot{y}'_1 = 0$. 

\bibliographystyle{unsrt}
\bibliography{biblio}

\end{document}